\documentstyle[prd,aps,epsfig,floats,axodraw]{revtex}
\begin{document}
\draft
\input epsf \renewcommand{\topfraction}{0.8} 
\newcommand{\beq}{\begin{equation}}
\newcommand{\eeq}{\end{equation}}
\newcommand{\pbar}{\not{\!\partial}}
\newcommand{\dbar}{\not{\!{\!D}}}
\def\lsim{\:\raisebox{-0.75ex}{$\stackrel{\textstyle<}{\sim}$}\:}
\def\gsim{\:\raisebox{-0.75ex}{$\stackrel{\textstyle>}{\sim}$}\:}
\twocolumn[\hsize\textwidth\columnwidth\hsize\csname 
@twocolumnfalse\endcsname
\title{Non-thermal leptogenesis with almost degenerate superheavy
neutrinos}    
\author{Rouzbeh Allahverdi$^{a}$ and Anupam Mazumdar$^b$}     
\address{$^a$Physik Department, TU M\"unchen, James Frank
Strasse, D-85748, Garching, Germany. \\
$^{b}$The Abdus Salam International Center for Theoretical 
Physics,~I-34100~Trieste, Italy.}
\date{\today} 
\maketitle
\begin{abstract}
We present a model with minimal assumptions for non-thermal leptogenesis 
with almost degenerate superheavy right-handed neutrinos in a supersymmetric 
set up. In this scenario the gauge singlet inflaton is directly coupled to 
the right-handed (s)neutrinos with a mass heavier than the inflaton mass. 
This helps avoiding potential problems which can naturally arise otherwise. 
The inflaton decays to the Standard Model leptons and Higgs via off-shell 
right-handed (s)neutrinos and reheats the Universe. The same channel is also 
responsible for generating the lepton asymmetry thus requiring no stage 
of preheating in order to excite superheavy (s)neutrinos. The suppressed
decay rate of the inflaton naturally leads to a sufficiently low reheat 
temperature, which in addition, prevents any wash out of the yielded 
asymmetry. We will particularly elaborate on important differences from 
leptogenesis with on-shell (s)neutrinos. It is shown that for nearly 
degenerate neutrinos a successful leptogenesis can be accommodated for 
a variety of inflationary models with a rather wide ranging inflationary 
scale. 
\end{abstract}


\vskip2pc]

\section{Introduction} 

The consistency of the abundance of the light elements synthesized
during the Big Bang nucleosynthesis (BBN) requires that the baryon 
asymmetry of the Universe (BAU), parameterized as 
$\eta_{\rm B}=(n_{\rm B}- n_{\bar{\rm B}})/s$ with $s$ being the entropy 
density and $n_{B}$ is the number density of the baryons to be in the 
range $(0.3-0.9)\times 10^{-10}$~\cite{bbn}. The asymmetry can be 
produced from the baryon symmetric Universe provided three conditions 
are simultaneously met; $B$ and/or $L$-violation, $C$- and $CP$-violation, 
and departure from thermal equilibrium~\cite{sakharov}. However any
produced asymmetry will be washed away by the SM $B+L$-violating sphaleron 
transitions which are active from temperatures $10^{12}$~GeV down to 
$100$~GeV~\cite{krs}, if $B-L=0$. Therefore, an asymmetry in $B-L$ is 
generally sought which is subsequently reprocessed in a thermal bath 
via sphalerons in order to yield a net baryon asymmetry given by 
$B=a(B-L)$. Here, $a$ is a model-dependent parameter; in case of 
the standard model (SM), $a = 28/79$, while in the minimal supersymmetric 
standard model (MSSM), $a =32/92$~\cite{khlebnikov}.

An attractive mechanism for producing $B-L$ asymmetry is from the
decay of the heavy right-handed (RH) Majorana neutrinos~\cite{fukugita}. 
Since the RH neutrinos are the SM singlets, a Majorana mass $M_N$, which
violates lepton number is compatible with all symmetries and hence can 
be arbitrarily large beyond the electroweak scale. This provides 
an elegant way for obtaining small masses $m_\nu$ for the light
neutrinos via the see-saw mechanism such that
$m_\nu \approx \left(m^{2}_{\rm D}/M_N \right)$~\cite{seesaw},
where $m_{\rm D}$ is the Dirac mass obtained from the Higgs vacuum 
expectation value (VEV). Moreover, a lepton asymmetry can be generated 
from the interference between the tree-level and the one-loop diagrams 
in an out-of-equilibrium decay of the RH neutrinos, provided $CP$-violating 
phases exist in the neutrino Yukawa couplings. The lepton asymmetry thus 
obtained will be partially converted to the baryon asymmetry
via sphaleron effects. This is the standard lore for producing lepton 
asymmetry commonly known as leptogenesis~\cite{fukugita,plumacher98}.

The present analyses of solar neutrino experiments favor the large
mixing angle MSW solution with 
$\Delta m^2_{\nu,solar}=6.1\times 10^{-5}~{\rm eV}^2$ and
$\tan^2\theta_{12}=0.41$~\cite{holanda}, while
$\Delta m^2_{\nu,atm}=3.2\times 10^{-3}~{\rm eV}^2$ and
$\sin^2(2\theta_{23})={(0.83-1)}$ provides the best fit to the
atmospheric neutrino data \cite{fogli01}.

In addition, cosmology \cite{dolgov02}, and neutrino less 
double-beta decay experiments \cite{kalpdor02} provide an upper
limit for the light neutrino masses. The masses and mixing angles 
which are required to explain solar and atmospheric neutrino data can be
obtained in both scenarios with hierarchical, or quasi-degenerate
neutrinos. Note that the hierarchical spectrum for heavy neutrinos
strongly suggests a spectrum of light neutrinos which is hierarchical
too unless there is a big conspiracy. On the other hand, a mild
hierarchy of RH neutrino masses could be compatible with the degenerate light
neutrinos with a certain amount of fine-tuning. In the former case,
one may consider thermal leptogenesis scenario where heavy neutrinos 
come into equilibrium with the primordial thermal bath through Yukawa 
interactions. The decay of the lightest RH neutrino easily satisfies 
the out-of-equilibrium condition by virtue of having a sufficiently 
small Yukawa coupling~\cite{plumacher98}. In a model-independent
analysis in Ref.~\cite{buchmuller02}, the authors have parameterized
thermal leptogenesis by four parameters; the $CP$ asymmetry, the heavy 
RH neutrino mass, the effective light neutrino mass, and the quadratic 
mean of the light neutrino masses. The final result was that an acceptable 
lepton asymmetry can be generated with 
$T_{\rm R} \sim M_{1}={\cal O}(10^{10})$~GeV, and
$\sum_{i}m_{\nu,i}< \sqrt{3}$~eV.

However the temperature required for thermal leptogenesis is
marginally compatible with the maximum allowed one in supersymmetric
theories, which is usually constrained by thermal gravitino production 
\cite{ellis,subir}. Gravitinos with a mass ${\cal O}(\rm TeV)$ decay 
much after nucleosynthesis and their decay products can change abundance 
of the light elements synthesized during BBN. For 
$100~{\rm GeV}\leq m_{3/2}\leq 1~{\rm TeV}$, a successful nucleosynthesis 
requires $n_{3/2}/s \leq (10^{-14}-10^{-12})$, which translates into 
$T_{\rm R}\leq (10^{7}-10^{10})~{\rm GeV}$ \cite{ellis,subir} 
\footnote{Recently, non-thermal production of helicity $\pm 3/2$
\cite{maroto}, and helicity $\pm 1/2$ gravitinos
\cite{kallosh,bastero-gil00} from inflaton oscillations have been 
considered. For a single chiral multiplet the helicity $\pm 1/2$ gravitino 
is the superpartner of the inflaton known as inflatino. The decay channels 
of inflatino have been discussed in Ref.~\cite{rouzbeh}. Also, it has been 
suggested \cite{rouzbeh}, and explicitly shown~\cite{nps}, that in 
realistic models with two chiral multiplets the helicity $\pm 1/2$ 
gravitino production is not a problem, so long as the inflationary
scale is sufficiently higher than the scale of supersymmetry breaking
in the hidden sector and the two sectors are weakly coupled. Gravitinos can
also be produced directly from the inflaton decay~\cite{nop}, and from 
the decay of the heavy stable neutral particles~\cite{allahverdi01}.}. 
The possible ways for obtaining a naturally low reheat temperature 
include gravitationally suppressed decay of the inflaton in models 
of high scale inflation~\cite{rs}, low scale inflationary
models~\cite{randall96,anu,grs}, a brief period of late thermal 
inflation~\cite{lyth95}, or, a completely new paradigm ``reheating 
through the surface evaporation'' which works even for high scale 
inflationary models~\cite{enqvist02}.

When the light neutrinos are almost degenerate 
$m_{\nu,1}\approx m_{\nu,2}\approx m_{\nu,3}$, which requires 
quasi-degenerate heavy neutrinos, the out-of-equilibrium condition in
thermal leptogenesis scenario cannot be satisfied in the minimal 
see-saw model~\cite{plumacher98}. More complicated models are required 
in this case \cite{compl}. On the other hand, if the mass splitting of 
the RH neutrinos becomes less than their decay widths, the perturbative 
calculations obviously break down. Then, the effect of finite decay 
widths of the RH neutrinos must be taken into account~\cite{pilaftsis}. 
The careful treatment of Ref.~\cite{pilaftsis} shows that a resonant 
enhancement of lepton asymmetry occurs in this case, while as expected, 
it vanishes in the limit of exactly degenerate neutrinos. This effect 
can be utilized to bring down the scale of heavy neutrino masses, and 
hence the leptogenesis scale~\cite{hambye01}.

However for almost degenerate heavy neutrinos, i.e. where the mass
splitting is larger than the decay width, one has to seek non-thermal 
leptogenesis (which works for the hierarchical neutrino masses as
well) in the minimal models. In this scenario RH neutrinos are produced 
non-thermally from the inflaton decay. This can occur during reheating 
if the inflaton decays to the RH neutrinos, which are lighter than the 
inflaton, with a considerable branching ratio \cite{kumekawa94}. Heavy 
neutrinos can also be produced via preheating~\cite{giudice99} 
(a stage of reheating where a resonant production of massive and/or 
massless bosons and fermions takes place \cite{preheating}), or, 
tachyonic preheating~\cite{tachyon}, even if the mass of boson/fermion 
exceeds that of the inflaton. All these are rather model-dependent and 
their main features can significantly vary from model to model. This is 
the prime reason why we do not pursue leptogenesis via preheating 
mechanism here.

In supersymmetric models one has the RH sneutrinos in addition. The 
sneutrinos are produced along with neutrinos during reheating, and 
with much higher abundances in preheating, thus serving as an additional 
source for leptogenesis~\cite{cdo}. Moreover, the RH sneutrinos can 
acquire a large VEV during inflation if their mass is less than the 
Hubble expansion rate during inflation $H_{I}$. Such a condensate starts 
oscillating once $H(t)\simeq M_N$, therefore automatically
satisfying the out-of-equilibrium condition. The decay of the sneutrino 
condensate can then yield the desired lepton asymmetry in the same 
fashion as neutrino decay~\cite{yanagida}, or via Affleck-Dine 
mechanism~\cite{berezhiani01}. This last scenario has an additional 
advantage that it solves the fine-tuning problem in $F$-term hybrid 
inflationary model in a very natural way~\cite{zurab}.

The success of all these scenarios, but preheating and the
Affleck-Dine oriented model, requires that the inflaton is heavier 
than the RH (s)neutrinos (in the hierarchical case inflaton 
only needs to be heavier than the lightest RH (s)neutrino.). 
Moreover, all the above scenarios are based upon the decay processes. An
attractive proposal was recently made, where the lepton asymmetry in
the visible sector is generated from the RH neutrino-mediated scattering
of the SM Higgs and leptons into a depleted hidden sector~\cite{bento}, 
rather than the decay of the on-shell heavy neutrinos.

In this paper we propose a simple supersymmetric model for non-thermal 
leptogenesis {\it without} any need of preheating mechanism. In this model
the inflaton is directly coupled to nearly degenerate RH (s)neutrinos 
which are heavier than the inflaton. Then the inflaton decay to the SM
fields, via off-shell RH (s)neutrinos, reheats the Universe and naturally 
leads to a sufficiently low reheat temperature. This same channel is also 
responsible for producing the lepton asymmetry.

In the next section we introduce our model and highlight several of 
its advantages. Then we turn to reheating and generation of the 
lepton asymmetry in this model and present our main results. In 
particular, we point out marked differences from leptogenesis with on-shell
(s)neutrinos. Finally, we conclude the paper with a brief summary.


\section{The model}

We start by introducing our model in a supersymmetric set up. The
relevant part of the superpotential is given by
\beq 
\label{super} 
W \supset {1 \over 2} m_\phi \Phi \Phi + {1 \over 2} g \Phi {\bf N}{\bf
N} + h{\bf N} {\bf H}_u {\bf L} + {1 \over 2} M_N {\bf N} {\bf N}\,.
\eeq 
Here $\Phi$, ${\bf N}$, ${\bf L}$, and ${\bf H}$ stand for the
inflaton, the RH neutrino, the lepton doublet, and the Higgs 
(which gives mass to the top quark) superfields, respectively.
Also, $m_\phi$ and $M_N$ denote inflaton and RH (s)neutrino masses,
respectively\footnote{Actually, $m_\phi$ denotes the frequency of the
inflaton oscillations around the global minimum of the potential. In
models of chaotic inflation $m_\phi \simeq H_I$. While in new and 
hybrid inflationary models it is usually (much) larger than $H_I$
\cite{linde}.}. We assume that the inflaton is coupled to the RH (s)neutrinos
via Yukawa coupling $g$, and $h$ denotes a typical neutrino Yukawa coupling. 
For simplicity, we have omitted all indices in $h$ matrix and 
superfields, and work in the basis where the Majorana mass matrix is 
diagonal. Further 
simplifications can be made for almost degenerate RH (s)neutrinos where 
$M_N$ is essentially the same for all of them. It is also 
conceivable in this case that the inflaton is coupled with the same
strength to three RH (s)neutrinos. This is particularly true when the 
inflaton has a nonzero VEV at the minimum which provides masses to the RH
(s)neutrinos. We focus on superheavy RH (s)neutrinos, i.e. assuming 
that $M_N \gg m_\phi$

Now let us discuss the merits why we seek RH (s)neutrinos heavier than 
the inflaton. If $M_N < m_\phi$, then one can easily produce on-shell
(s)neutrinos from the inflaton decay, either perturbatively or via 
preheating. First consider (s)neutrino production
in perturbative inflaton decay. A perturbative decay requires a small
coupling to the (s)neutrinos. This is naturally achieved when the inflaton 
lies in a hidden sector which is only gravitationally coupled to the SM 
sector~\cite{rs}. In this case the total decay rate of the 
inflaton is given by $\Gamma_{\rm d} \sim m^{3}_{\phi}/M^{2}_{\rm P}$, 
while the partial decay rate to (s)neutrinos is given by
$\Gamma_{\phi \rightarrow N} \sim m_\phi M^{2}_{N}/M^{2}_{\rm P}$
\cite{allahverdi01}, where $M_{\rm P} = 2.4 \times 10^{18}$~GeV 
is the reduced Planck mass. This results in a branching ratio 
$\simeq \left(M_N/m_\phi \right)^2$, which is too small if 
$M_N \ll m_\phi$. Note that a successful leptogenesis requires an
acceptable branching ratio given the entropy generation from reheating, thus
implying that $M_N$ must not be much smaller than $m_\phi$.

Besides, a small coupling $g$, which is required to ensure a perturbative 
treatment, leads to another potential problem. The sneutrino field 
$\tilde N$ can acquire a large VEV during inflation. Since $\tilde N$ 
is directly coupled to the inflaton, it might even ruin the
flatness of the inflaton potential. On a lighter note, 
$\langle \tilde N \rangle$ remains non-vanishing after the end of 
inflation in any case and may contribute to isocurvature density 
perturbations~\cite{liddle00}. This requires a delicate treatment 
of a coupled system which depends on the choice of a model. This 
is an issue which has been side-lined in most supersymmetric models 
of non-thermal leptogenesis, except Ref.~\cite{berezhiani01}.

If $g$ is sufficiently large, (s)neutrinos may be produced in a 
non-perturbative manner during the stage of (tachionic) 
preheating~\cite{giudice99,tachyon}. For the superpotential in 
Eq.~(\ref{super}), the necessary condition for preheating reads 
$g \phi_0 > m_\phi$, where $\phi_0$ is the initial
amplitude of the inflaton oscillations. This guarantees that $\tilde N$ 
is heavier than the inflaton during inflation, and hence 
$\langle \tilde N \rangle = 0$ after the end of inflation, resulting 
in a simpler initial condition in post-inflationary era. On the other 
hand, both RH neutrinos and sneutrinos can be produced via preheating 
(sneutrinos much more abundantly by virtue of obeying the Bose 
statistics~\cite{preheating}). However, as mentioned earlier, this 
is rather model-dependent. For example, if the inflaton has a VEV 
at the minimum, denoted as $v$, then it is hard to envisage an 
efficient production of (s)neutrinos through parametric resonance. 
The reason is that $M_N = gv$ and $\phi_0 \simeq v$ in this
case, which implies $g \phi_0 \simeq M_N$. It is therefore evident that 
there will be no preheating of (s)neutrinos for $M_N < m_\phi$. On the 
other hand, for $M_N > m_\phi$ preheating is possible only if 
$g \phi_0 m_\phi \gg M^{2}_{N}$~\cite{preheating,kt1}. In particular, 
resonant creation rapidly ceases to be efficient for $M_N > 10 m_\phi$ 
\cite{kt1}\footnote{It has been shown in Ref.~\cite{kt1}, that for a 
quadratic potential; $V_\phi \sim m^{2}_{\phi} \phi^2$, efficient
resonant production of particles with a mass $M_N = 10 m_\phi$
requires $g \phi_0 > 10^{4} m_\phi$. On the other hand, for a
quartic potential $V_\phi \sim \lambda \phi^4$, preheating
of these particles practically disappears. Preheating in supersymmetric
hybrid inflation model is also not efficient~\cite{heitmann}.}. In the
tachyonic preheating scenario, too, the produced (s)neutrinos usually
have an abundance much less than the inflaton abundance when 
$M_N \gg m_\phi$ \cite{tachyon}. In conclusion, it is very difficult (if not
impossible) to obtain the desired lepton asymmetry in a wide range of 
inflationary models, by solely relying on non-perturbative dynamics.

Now we count upon the advantages of our model. First of all, for 
$M_N \gg m_\phi$ the post-inflationary dynamics is simpler since
$\langle\tilde N\rangle = 0$ at the end of inflation. The Universe is
reheated through the inflaton decay to the Higgs and SM leptons via 
off-shell RH (s)neutrino. The decay rate, as we will see shortly, is 
suppressed as $\left(m_\phi/M_N \right)^4$. This naturally leads to 
an acceptably low reheat temperature when $M_N \gg m_\phi$. Furthermore, 
the inflaton decay alone is responsible for the generation of the 
lepton asymmetry. This makes the model minimal since leptogenesis is 
now directly connected with reheating. Also, the washing out of the lepton 
asymmetry from thermal scattering of the SM leptons and Higgs is 
completely negligible since $T_{\rm R} \ll M_N$.

Our main focus will be on almost degenerate light neutrinos, which 
can be derived naturally from almost degenerate RH neutrinos. 
An example of such a model is presented in Ref.~\cite{fhy}, where 
neutrino masses and mixing compatible with the solar and atmospheric 
neutrino solutions are derived in the framework of democratic mass 
matrix. There the neutrino Yukawa matrix ${\bf h}$ is almost diagonal in 
the same basis as the Majorana mass matrix. This makes sense since when both 
are proportional to the identity matrix the light 
neutrinos come out to be exactly degenerate. Then by perturbing
around this pattern, we can obtain a nearly degenerate
texture. In the calculations below, $M_N$ and $\Delta M_N$ denote the nearly
equal diagonal elements of the Majorana mass matrix and their typical
differences respectively. Also $h$ and $\delta h$ represent the nearly 
equal diagonal elements of the Yukawa matrix and their differences 
respectively, while $h'$ stands for the typical non-diagonal elements. 
It is assumed that $\Delta M_N < M_N$ and $h'< \delta h < h$.

\section{Reheating the Universe}

The main decay mode of the inflaton is to a four-body final state
consisting of two Higgs/Higgsino-lepton/slepton particles (and their 
$CP$ transforms). Since, we have assumed $m_\phi \ll M_N$, it is essential 
to find those diagrams which are least suppressed by powers of $M_N$. 
These diagrams, shown in Fig.~(1), which arise from the leading order terms in
the effective superpotential after integrating out ${\bf N}$,
is given by
\beq \label{effective}
W_{eff} \supset {1\over 2} m_\phi \Phi^2 + {1 \over 2 M^{2}_{N}} g
h^2 \Phi ({\bf H}_u {\bf L})({\bf H}_u {\bf L}). 
\eeq
We should therefore choose that part of the $N$ propagator with a mass
insertion, namely the part suppressed as $1/M_N$ (the other part of 
the propagator is proportional to $m_{\phi}/M^{2}_N$). In diagrams below 
two opposite arrows on the $N$ propagator represent this dominant part. 
Note that $\tilde N$ propagator is proportional to $\left(1/M^{2}_{N}\right)$ 
to the leading order.

First, we evaluate the rate for inflaton decay without any specific 
assumption about Majorana masses and Yukawa couplings (except that $g$ is 
diagonal and universal, and $m_\phi \ll M_N$). Generically, the trajectory 
of the inflaton motion is a line on a complex $\phi$ plane. We can 
therefore assume, without loss of generality, that only the real component 
of the inflaton has a VEV, thus treating the decaying inflatons as real 
fields. In addition, the SM particles are much lighter than the inflaton 
in the case under consideration (as will be confirmed by our results). 
Then the phase space factor for the four-body decay is readily found 
to be $\left[16 \cdot 96 \cdot (2\pi)^5\right]^{-1}$. The inflaton 
coupling to a given final state consisting of ${\bar L}_j$ (or 
${\bar {\tilde L}}_j$) and ${\bar L}_k$ (or ${\bar {\tilde L}}_k$), 
plus two ${\bar H}_u$ (or ${\bar {\tilde H}}_u$), is given by 
$\sum_{i} {g h_{ij} h_{ik} \over 2 M^{2}_{i}}$. Here $j$ and $k$ stand for 
the lepton flavor. There is also a multiplicity factor for each final 
state which can be calculated easily.

Given all possible weak isospin assignments, with flavour indices fixed, 
there exist a total of nine final states. Seven of them which consist of two 
fermions and two scalars are
{\it (1) ${\bar L}^{a}_{j}{\bar L}^{a}_{k}{\bar H}^{b}_{u}{\bar H}^{b}_{u}$}, 
{\it (2) ${\bar{\tilde L}}^{a}_{j}{\bar
{\tilde L}}^{a}_{k}{\bar{\tilde H}}^{b}_{u}{\bar{\tilde H}}^{b}_{u}$}, 
{\it (3)${\bar L}^{a}_{j}{\bar L}^{b}_{k}{\bar H}^{a}_{u}{\bar H}^{b}_{u}$}, 
{\it (4) ${\bar{\tilde L}}^{a}_{j}{\bar{\tilde L}}^{b}_{k}{\bar{\tilde H}}^{a}_
{u}{\bar{\tilde H}}^{b}_{u}$}, {\it (5) ${\bar L}^{a}_{j}{\bar{
\tilde L}}^{a}_{k}{\bar H}^{b}_{u}{\bar{\tilde H}}^{b}_{u}$}, 
{\it (6) ${\bar L}^{a}_{j}{\bar{\tilde L}}^{b}_{k}{\bar H}^{a}_{u}{\bar
{\tilde H}}^{b}_{u}$}, and {\it (7) ${\bar L}^{a}_{j}{\bar{\tilde L}}^{b}_{k} 
{\bar H}^{b}_{u} {\bar{\tilde H}}^{a}_{u}$}. There are also two 
final states consisting of four scalars: {\it (8) 
${\bar{\tilde L}}^{a}_{j}{\bar{\tilde L}}^{a}_{k}{\bar H}^{b}_{u} 
{\bar H}^{b}_{u}$} and 
\begin{center}
\SetScale{0.6} \SetOffset(40,55)
\begin{picture}(175,160)(0,0)
\DashLine(0,80)(75,80){5} \Text(0,40)[l]{$\phi$}
\Vertex(75,80){3}
\ArrowLine(100,55)(125,30)
\ArrowLine(100,55)(75,80) \Text(60,70)[h]{$N$}
\ArrowLine(100,105)(125,130)
\ArrowLine(100,105)(75,80) \Text(60,20)[b]{$N$}
\Vertex(125,130){3}
\ArrowLine(175,160)(125,130) \Text(135,100)[r]{${\bar L}({\bar{\tilde
H}}_{u})$}
\DashArrowLine(175,100)(125,130){5} \Text(135,60)[r]{${\bar H}_u(\bar
{\tilde L})$}
\Vertex(125,30){3}
\DashArrowLine(175,60)(125,30){5} \Text(135,40)[r]{${\bar H}_u(\bar
{\tilde L})$}
\ArrowLine(175,0)(125,30) \Text(135,0)[r]{${\bar L}({\bar{\tilde H}}_{u})$}
\end{picture}
\vspace*{-13mm}

{\large a)~$\Delta L = -2$}

\end{center}

\begin{center}
\SetScale{0.6} \SetOffset(40,55)
\begin{picture}(175,160)(0,0)
\DashLine(0,80)(75,80){5} \Text(0,40)[l]{$\phi$}
\Vertex(75,80){3}
\DashArrowLine(75,80)(125,30){5} \Text(60,70)[h]{$\tilde N$}
\DashArrowLine(125,130)(75,80){5} \Text(60,20)[b]{$\tilde N$}
\Vertex(125,130){3}
\DashArrowLine(175,160)(125,130){5} \Text(135,100)[r]{${\bar H}_{u}$}
\DashArrowLine(175,100)(125,130){5} \Text(135,60)[r]{$\bar {\tilde L}$}
\Vertex(125,30){3}
\ArrowLine(175,60)(125,30) \Text(135,40)[r]{${\bar{\tilde H}}_{u}$}
\ArrowLine(175,0)(125,30) \Text(135,0)[r]{${\bar L}$}       
\end{picture}
\vspace*{-13mm}

{\large b)~$\Delta L = -2$}

\end{center}

\begin{center}
\SetScale{0.6} \SetOffset(40,55)
\begin{picture}(175,160)(0,0)
\DashLine(0,80)(75,80){5} \Text(0,40)[l]{$\phi$}
\Vertex(75,80){3}
\DashArrowLine(125,30)(75,80){5} \Text(60,70)[h]{$\tilde N$}
\DashArrowLine(125,130)(75,80){5} \Text(60,20)[b]{$\tilde N$}
\Vertex(125,130){3}
\DashArrowLine(175,160)(125,130){5} \Text(135,100)[r]{${\bar H}_{u}$}
\DashArrowLine(175,100)(125,130){5} \Text(135,60)[r]{$\bar {\tilde L}$}
\Vertex(125,30){3}
\DashArrowLine(175,60)(125,30){5} \Text(135,40)[r]{${\bar H}_{u}$}
\DashArrowLine(175,0)(125,30){5} \Text(135,0)[r]{$\bar {\tilde L}$}    
\end{picture}
\vspace*{-10mm}

{\large c)~$\Delta L = -2$}

\end{center}

\vspace*{7mm}

\noindent
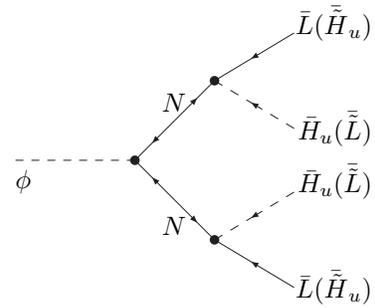
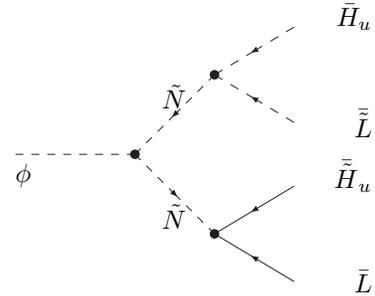
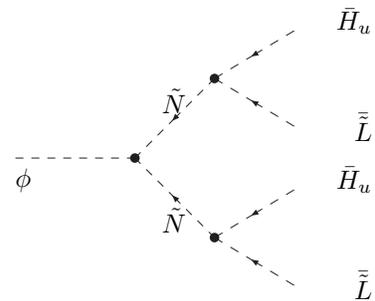
{\bf Fig. 1:}~Diagrams together with their $CP$ transformed, for which 
$\Delta L = +2$, represent the inflaton decay into two 
Higgs/Higgsino-lepton/slepton pairs at the leading order.     
\vspace*{3mm}
\newline
{\it (9) ${\bar{\tilde L}}^{a}_{j} {\bar
{\tilde L}}^{b}_{k} {\bar H}^{a}_{u} {\bar H}^{b}_{u}$}.  

Note that at each vertex in diagram (a), the production of
${\bar L}^{a}$ (or ${\bar{\tilde L}}^{a}$) is accompanied by that of 
${\bar H}^{b}_u$ (or ${\bar{\tilde H}}^{b}_{u}$), and vice versa. In 
diagram (b), on the other hand, the production of
${\bar L}^{a}$ (or ${\bar{\tilde L}}^{a}$) is accompanied by that of 
${\bar{\tilde H}}^{b}_{u}$ (or ${\bar H}^{b}_u$), and vice versa. This 
implies that final states in $\it {(1)-(4)}$ and $\it {(7)}$ can arise 
from diagram (a), while $\it {(6)}$ arises only from diagram (b). On the 
other hand, the final state in $\it {(5)}$ can arise from both diagrams. 
Finally, $\it {(8),(9)}$ arise only from diagram (c).

The rate for the inflaton decay to the final states in 
${\it (1), (2)}$ and ${\it (8)}$ are the same and given by 
\begin{eqnarray} 
\label{partial1}
\Gamma_{1} &= &\Gamma_{2}=\Gamma_{8}\, \nonumber \\ 
& \simeq & \sum_{j \leq k} \left(2 \cdot 
[8 - 4 \delta_{jk}]\right) \times {m^{5}_{\phi} \over 16 \cdot 96 \cdot 
{(2 \pi)}^5}{\left |\sum_{i} g {h_{ij} h_{ik} \over 2 M^{2}_{i}}\right |}^{2}.
\end{eqnarray} 
The constraint $j\leq k$ is imposed in order to avoid double-counting 
of the same final states. Note that the first number inside parenthesis 
comes from the summation over all isospin states, while the second one 
represents the overall factor from the superposition of different 
contributions for each isospin assignment.

Similarly, one can also evaluate the rates for the decay into other final 
states. The results are
\beq 
\label{partial3}
\Gamma_{3}=\Gamma_{4}=\Gamma_{9}={1 \over 2} \Gamma_{1}.      
\eeq   
While 
\begin{eqnarray}
\label{partial5}
\Gamma_{6} &= &\Gamma_{7}={1 \over 4} \Gamma_{5}\, \nonumber \\ 
& \simeq & \sum_{j,k} \left( 2 \cdot 4 \right) \times {m^{5}_{\phi} 
\over 16 \cdot 96 \cdot {(2 \pi)}^5}{\left|\sum_{i} g {h_{ij} h_{ik} 
\over 2 M^{2}_{i}}\right|}^{2}.      
\end{eqnarray}   
The total decay rate of the inflaton will be
\beq \label{total}
\Gamma_{\rm d} = \sum^{9}_{i=1} \Gamma_i.
\eeq
Let us come back to the case with nearly degenerate neutrinos, where 
$\Delta M_N < M_N$ and $h'<\delta h < h$. In this case each 
$N_i$ $({\tilde N}_i)$ is dominantly coupled to the $i$-th lepton 
doublet, and the coupling is $h$. In consequence, off-shell 
$N_i$ $({\tilde N}_i)$ mainly contributes to the inflaton decay to 
the final states with $j = k = i$. Then we can show that the decay 
rate will be given by 
\beq 
\label{decay}
\Gamma_{\rm d} \simeq {21 \over 2^{14} \pi^5} g^2 h^4 {m^{5}_{\phi} \over
M^{4}_{N}}\,. 
\eeq

The inflaton completely decays when $H \simeq \Gamma_{\rm d}$, where
$H \simeq ( g^{1/2}_{*} T^2/M_{\rm P})$ in a radiation-dominated 
Universe \cite{linde}, with $g_*$ being the effective number of
relativistic degrees of freedom which is $\simeq 214$ in the MSSM. 
Assuming that thermal equilibrium is achieved when 
$H \simeq \Gamma_{\rm d}$ (which is justifiable; for the detailed 
discussion, see Refs.~\cite{ds,ad2}), we obtain
\beq \label{rehtemp}
{T_{\rm R} \over m_\phi} \simeq 10^{-7/2} {g h^2 m^{3/2}_{\phi} M^{1/2}_{\rm
P} \over M^{2}_{N}}\,. 
\eeq            

Some comments are in order regarding our estimates of $\Gamma_{\rm d}$
and $T_{\rm R}$. One might think that inflaton decaying into four
scalars, the same as in diagram (b) except that $\bar {{\tilde H}_u}$
and $\bar L$ are replaced with $H_u$ and $\tilde L$, would occur at a
rate only suppressed by two powers of $M_N$. However, this is
not the case since this leading order contribution is canceled out
by that from another diagram and the overall rate is actually
proportional to $(m^{7}_{\phi}/M^{6}_{N})$. This is just the 
manifestation that these diagrams do not arise from the effective
superpotential individually. Also, there exists a two-body decay
channel for the inflaton, into 
${\bar H}_u H_u ({\bar{\tilde H}}_u {\tilde H}_u)$ or 
${\bar L} L ({\bar {\tilde L}} {\tilde L})$, at the one-loop 
level. It can easily be derived by choosing
$\left(1/M_N\right)$ and $\left(m_{\phi}/M^{2}_{N}\right)$ parts of
$N$ propagators in diagram (a) and connecting the 
${\bar H}_u(\bar{\tilde L})$, or ${\bar L}({\bar{\tilde H}}_u)$, lines. 
This channel has a much larger phase space factor ${(8 \pi)}^{-1}$, 
while the dependence on $g$ and $h$ remains the same as in Fig. (1). 
However, the two-body decay rate is $\propto (m^{7}_{\phi}/M^{6}_{N})$. 
Thus by taking the one-loop factor ${(4 \pi)}^{-2}$ into account and 
for $M_N \geq 10m_\phi$, it will eventually be smaller than that in 
Eq.~(\ref{decay}).

Finally, the inflaton can also decay into the SM fields via gravitational 
couplings with a decay rate 
$\Gamma_{grav}\sim (v/M_{\rm P})^2(m^{3}_{\phi}/M^{2}_{\rm P})$, 
where $v$ denotes inflaton VEV at the global minimum of the
potential~\cite{allahverdi01}. Such a decay rate can however be 
neglected compared to the four-body decay provided $v \ll M_{\rm P}$.

\section{The lepton asymmetry}

In this section we evaluate the lepton asymmetry generated from the 
inflaton decay through diagrams in Fig.~(1). First, we remind the readers 
that for the standard case where the decay of on-shell neutrinos yields 
the lepton symmetry, one has $\eta_{\rm L}=\Sigma_i \epsilon_i(n_{N_i}/s)$, 
where
\beq \label{asymmetry}
\epsilon_i = \sum_{i \neq j} \epsilon_{ij}~;~\epsilon_{ij} = - {1 \over 8
\pi} {1 \over [{\bf h} {\bf h}^{\dagger}]_{ii}} {\rm Im}
\left( [{\bf h} {\bf h}^{\dagger}]_{ij}\right)^2 f \left({M^{2}_{j}
\over M^{2}_{i}} \right)\,,
\eeq          
and \cite{one-loop}
\beq \label{corrections}
f(x) = \sqrt{x} \left ({2 \over x-1} + {\ln} \left[{1 + x \over
x}\right ] \right ).
\eeq
The first and second terms on the right-hand side of Eq.~(\ref{corrections})
correspond to the one-loop self-energy, and vertex corrections, respectively. 
For hierarchical $N$, the following lower bound is found \cite{di} 
\beq \label{hierarchical}
|\epsilon_1| \leq {3 \over 8 \pi} {M_1 m_3 \over {\langle H_0 \rangle}^{2}},
\eeq
where $M_1$ and $m_3$ denote masses of the lightest heavy neutrino and the 
heaviest light neutrino respectively. Here 
$\langle H^{0}_{u}\rangle = 174~{\rm sin}\beta$~GeV is the VEV 
of $H_u$ in our vacuum, with ${\rm tan} \beta$ defined as the ratio of
$\langle H^{0}_{u} \rangle$ and $\langle H^{0}_{d} \rangle$. 
On the other hand, $x\approx 1$ for almost degenerate RH neutrinos, and 
hence the self-energy contribution dominates. Then, it can be shown that (to
the leading order) \cite{fhy}
\beq 
\label{epsilon}
\epsilon_1 \simeq \epsilon_2 \simeq \epsilon_3 \simeq {1
\over 4 \pi} {h'^{2} \over h^2} {M_N \over \Delta M_N}
{M_N \over {\langle H^{0}_{u}
\rangle}^2} {\Delta m^{2}_{\nu,atm} \over 2 m_{\nu}}\,.
\eeq

Now we come back to our case, where the inflaton decay via 
off-shell $N$ $({\tilde N})$ produces the lepton asymmetry. 
The net lepton asymmetry is generated from the interference 
between diagrams in Fig (1), and the one-loop diagrams
representing self-energy and vertex corrections to one of the $N$
$(\tilde N)$ propagators. Diagrams with one-loop correction to 
both $N$ $(\tilde N)$ legs are of higher order and will be subdominant. There 
are major differences which arise in the analysis compared to the on-shell 
case, as we note in this following discussion. To demonstrate these 
differences explicitly, we focus on self-energy and vertex corrections 
to the diagram (a) of Fig. (1), shown in Fig. (2). Similar 
arguments will go through for the inflaton decay through 
diagrams (b) and (c) in Fig. (1).

Note that both $H_u L$ and ${\tilde H}_u {\tilde L}$ loops contribute 
to the self-energy correction, while only one of them is relevant 
in the vertex correction for a given final state. Also, recall that 
only loops with on-shell particles make a contribution to the 
resultant asymmetry. Thus the self-energy and vertex loops involving 
$N_l$ actually represent s-channel and t-channel scattering of 
a Higgs-lepton or Higgsino-slepton pair via off-shell $N_l$, respectively. 
The center-of-mass energy available in these processes is at most equal to the 
inflaton mass. In consequence, the self-energy correction is simply twice as 
large as the vertex correction for $m_\phi \ll M_N$ 
\footnote{This is similar to the $x \gg 1$ limit for the standard case 
in Eq.~(\ref{corrections})}. It can also be shown that only the mass 
insertion part of $N_l$ propagator contributes to the generated asymmetry from 
self-energy correction of $N_i$. The diagram with mass insertion in $N_i$ 
propagator will be irrelevant, exactly like the standard case \cite{one-loop}.

Important difference arises in comparison with the standard case 
is that there the center-of-mass energy in the two-body decay 
of $N_i$ is simply determined by $M_i$. While here 
the energy flowing in the $N_i$ leg is $0 < E < m_\phi$. In the 
$m_\phi \ll M_N$ limit, the $N_i$ propagator is $E/M^{2}_{i}$, 
while the $N_l$ propagator will simply be $1/M_l$, see diagrams in 
Fig. (2). For a given final state with definite momenta the 
one-loop diagram is suppressed as $E^{2}/M_i M_l$ with respect to 
the tree-level one. Upon performing the phase space integration 
over a four-body final state we find the suppression will be 
$m^{2}_{\phi}/M_i M_l$ times some numerical factor $\sim {\cal O}(1)$ . 
For 

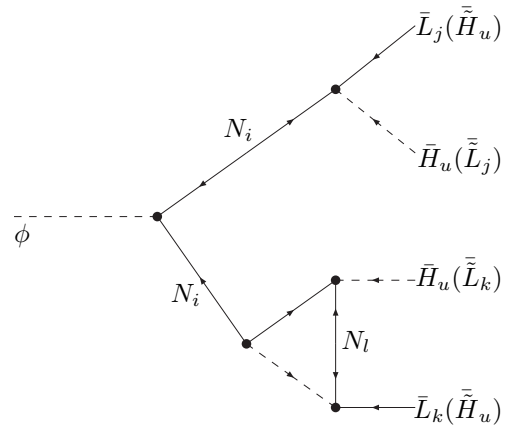
\begin{center}
\SetScale{0.6} \SetOffset(40,95)
\begin{picture}(252,240)(0,0)
\DashLine(0,120)(90,120){5} \Text(0,65)[l]{$\phi$}
\Vertex(90,120){3}
\ArrowLine(188,50)(174,60) \Text(108,20)[b]{$N_l$}
\ArrowLine(188,50)(202,40)
\ArrowLine(118,100)(90,120) \Text(85,105)[h]{$N_i$}
\ArrowLine(118,100)(174,60)
\Vertex(174,60){3}
\Vertex(118,100){3}
\DashArrowArc(146,80)(34,146,-34){7}
\ArrowLine(146,160)(202,200)
\ArrowLine(146,160)(90,120) \Text(55,57)[b]{$N_i$}
\Vertex(202,200){3}
\ArrowLine(252,240)(202,200) \Text(185,145)[r]{${\bar L}_{j} ({\bar{\tilde
H}}_{u})$}
\DashArrowLine(252,160)(202,200){5} \Text(185,95)[r]{${\bar H}_u({\bar
{\tilde L}}_{j})$}
\Vertex(202,40){3}
\DashArrowLine(252,80)(202,40){5} \Text(185,55)[r]{${\bar H}_u({\bar
{\tilde L}}_{k})$}
\ArrowLine(252,0)(202,40) \Text(185,0)[r]{${\bar L}_{k} ({\bar 
{\tilde H}}_{u})$}
\end{picture}
\vspace*{-30mm}

\begin{center}
\SetScale{0.6} \SetOffset(40,95)
\begin{picture}(252,240)(0,0)
\DashLine(0,120)(90,120){5} \Text(0,65)[l]{$\phi$}
\Vertex(90,120){3}
\ArrowLine(188,50)(174,60) \Text(108,20)[b]{$N_l$}
\ArrowLine(188,50)(202,40)
\ArrowLine(118,100)(104,110) \Text(85,105)[h]{$N_i$}
\ArrowLine(90,120)(104,110)
\ArrowLine(118,100)(174,60)
\Vertex(174,60){3}
\Vertex(118,100){3}
\DashArrowArc(146,80)(34,146,-34){7}
\ArrowLine(90,120)(202,200) \Text(55,57)[b]{$N_i$}
\Vertex(202,200){3}
\ArrowLine(252,240)(202,200) \Text(185,145)[r]{${\bar L}_{j} ({\bar{\tilde
H}}_{u})$}
\DashArrowLine(252,160)(202,200){5} \Text(185,95)[r]{${\bar H}_u({\bar
{\tilde L}}_{j})$}
\Vertex(202,40){3}
\DashArrowLine(252,80)(202,40){5} \Text(185,55)[r]{${\bar H}_u({\bar
{\tilde L}}_{k})$}
\ArrowLine(252,0)(202,40) \Text(185,0)[r]{${\bar L}_{k} ({\bar 
{\tilde H}}_{u})$}
\end{picture}
\end{center}
\vspace*{-30mm}

{\large a)~}

\end{center}

\vspace{2mm}

\begin{center}
\SetScale{0.6} \SetOffset(40,95)
\begin{picture}(252,240)(0,0)
\DashLine(0,120)(90,120){5} \Text(0,65)[l]{$\phi$}
\Vertex(90,120){3}
\ArrowLine(146,160)(202,200)
\ArrowLine(146,160)(90,120) \Text(85,105)[h]{$N_i$}
\Vertex(202,200){3}
\ArrowLine(146,40)(90,120) \Text(65,40)[b]{$N_i$}
\ArrowLine(252,240)(202,200) \Text(185,145)[r]{${\bar L}_{j} ({\bar{\tilde
H}}_{u})$}
\DashArrowLine(252,160)(202,200){5} \Text(185,95)[r]{${\bar H}_u({\bar
{\tilde L}}_{j})$}
\Vertex(146,40){3}
\ArrowLine(146,40)(202,80)
\DashArrowLine(146,40)(202,0){5} 
\ArrowLine(202,40)(202,80)
\ArrowLine(202,40)(202,0) \Text(135,25)[r]{$N_l$}
\DashArrowLine(252,80)(202,80){5} \Text(185,50)[r]{${\bar H}_u({\bar
{\tilde L}}_{k})$}
\Vertex(202,80){3}
\ArrowLine(252,0)(202,0) \Text(185,0)[r]{${\bar L}_{k} ({\bar 
{\tilde H}}_{u})$}
\Vertex(202,0){3}
\end{picture}
\vspace*{-30mm}
\vspace{25mm}

\begin{center}
\SetScale{0.6} \SetOffset(40,95)
\begin{picture}(252,240)(0,0)
\DashLine(0,120)(90,120){5} \Text(0,65)[l]{$\phi$}
\Vertex(90,120){3}
\ArrowLine(90,120)(202,200) \Text(85,105)[h]{$N_i$}
\Vertex(202,200){3}
\ArrowLine(146,40)(118,80)
\ArrowLine(90,120)(118,80) \Text(65,40)[b]{$N_i$}
\ArrowLine(252,240)(202,200) \Text(185,145)[r]{${\bar L}_{j} ({\bar{\tilde
H}}_{u})$}
\DashArrowLine(252,160)(202,200){5} \Text(185,95)[r]{${\bar H}_u({\bar
{\tilde L}}_{j})$}
\Vertex(146,40){3}
\ArrowLine(146,40)(202,80)
\DashArrowLine(146,40)(202,0){5} 
\ArrowLine(202,40)(202,80)
\ArrowLine(202,40)(202,0) \Text(135,25)[r]{$N_l$}
\DashArrowLine(252,80)(202,80){5} \Text(185,50)[r]{${\bar H}_u({\bar
{\tilde L}}_{k})$}
\Vertex(202,80){3}
\ArrowLine(252,0)(202,0) \Text(185,0)[r]{${\bar L}_{k} ({\bar 
{\tilde H}}_{u})$}
\Vertex(202,0){3}
\end{picture}
\vspace*{-30mm}
\end{center}

{\large b)~}

\end{center}

\vspace*{7mm}

\noindent
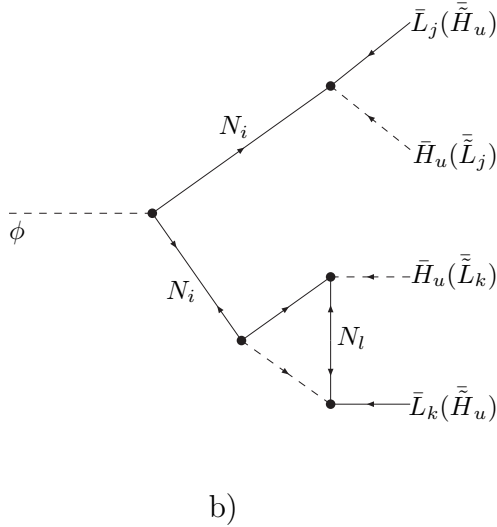
{\bf Fig. 2:}~Diagrams representing one-loop (a) self-energy, and 
(b) vertex corrections to the decay channel shown in Fig. (1a). The 
interference between these and the tree-level diagram results in a 
net lepton asymmetry.       
\vspace*{3mm}
\newline

simplicity, we take the average 
energy in the $N_i$ legs to be $m_\phi/2$, and hence the suppression 
comes as $\simeq m^{2}_{\phi}/4 M_i M_l$. This 
approximation is adequate for our purposes in the $m_\phi \ll M_N$ limit, and 
any difference from the exact result will be numerically irrelevant.
The reason is that the main contribution to the phase space integral comes 
from the bulk of the available phase space. While, contribution of the 
parts in which the energy of some decay products is $\ll m_\phi$, 
including parts with $E \approx 0$ or $E \approx m_\phi$, is suppressed. 
The situation will be more complicated when $m_\phi$ and $M_N$ are 
not very different, since the energy and momentum carried by $N$ legs 
is comparable to $M_N$. In such a case the $N_i$ and $N_l$ 
propagators can strongly depend on the phase space distribution of the decay 
products and the above approximation may not be sufficient.

Now let us find the asymmetry parameter in the inflaton decay. First consider 
diagrams in Figs. (1a) and (2). For a given final state
the tree-level and interference terms naturally have the same 
multiplicity factor. As explained earlier the self-energy correction 
is twice as large as the vertex correction, and also, the average energy 
carried by each of the $N$ legs can be approximately taken to be $m_\phi/2$. 
The contributions from both diagrams in Figs. (2a) and (2b) are equal, and 
hence the asymmetry receives an extra factor of 
$3 m^{2}_{\phi}/2 M^{2}_{N}$, in addition to $(1/8 \pi)$ prefactor in Eq. 
(\ref{asymmetry}). Note that the one-loop correction can come from each of 
the two $N$ legs, 
which is equivalent to exchanging $j$ with $k$. The situation will be similar 
for the asymmetry in 
the inflaton decay through diagrams (b) and (c) in Fig. (1). Thus, after 
summing over all possible final states, we obtain
\beq \label{asymmetry1}
\epsilon \simeq  - {3 \over 8 \pi} \times {\sum_{i,n,l} {{\rm Im} 
\left[({\bf h} {\bf h}^{\dagger})_{ni} 
({\bf h} {\bf h}^{\dagger})_{nl} ({\bf h} {\bf h}^{\dagger})_{il}\right]
m^{2}_{\phi}
\over M^{3}_{i} M^{2}_{n} M_{l}} \over \sum_{i,n} 
{([{\bf h} {\bf h}^{\dagger}]_{in})^{2} \over M^{2}_{i} M^{2}_{n}}},
\eeq                       
which is functionally very different from the standard case in Eq. 
(\ref{asymmetry}).

We now come back to the case with nearly degenerate neutrinos. Now, the 
denominator of the second term on the right-hand side of 
Eq. (\ref{asymmetry1}) is $\simeq 3 h^{4}/M^{4}_{N}$. In this case the Yukawa 
matrix ${\bf h}$ is almost diagonal, and so is the matrix ${\bf h} {\bf h}^
{\dagger}$. The 
numerator receives the main contribution from the terms with $i=n$ and $i=l$
\footnote{Terms with $n=l$ and $i=n=l$ are real, and hence do not contribute 
to the asymmetry}, and can be written as    
\beq 
\label{ndc}
\sum_{i \neq l} h^{2} {\rm Im} \left([{\bf h} {\bf h}^{\dagger}]_{il} 
\right)^{2} ~ m^{2}_{\phi} \left({1 \over M^{5}_{i} M_{l}} - {1 \over 
M^{4}_{i} M^{2}_{l}} \right).
\eeq      
This can further be simplified to 
$12 h^{3} {\delta h} h'^{2} (m^{2}_{\phi}{\Delta M}_N/M^{7}_{N})$, assuming 
that only the non-diagonal elements contain $CP$-violating phases. 
Since the lepton number is violated by two units in the inflaton decay, 
we finally have
\beq 
\label{finallepton}
{n_{L} \over n_{\phi}} \simeq {3 \over \pi} {\delta h h'^{2} \over h} 
{{\Delta M}_{N} \over M_{N}} \left({m_\phi \over M_N}\right)^{2}.
\eeq    
An important observation is that here the final asymmetry is proportional to 
$\Delta M_N$, contrary to the on-shell case in Eq. (\ref{epsilon}). 
Therefore the generated asymmetry actually decreases as the RH (s)neutrinos 
become more degenerate. This is not difficult to understand as the available 
energy in the inflaton decay $m_\phi$ is far below the mass of the RH 
(s)neutrinos $M_N$, independently of how degenerate the latter ones are.

The total asymmetry in the baryons (after taking into account of sphaleron
effects) can be expressed as  
\begin{eqnarray} 
\label{final}
\eta_{\rm B} &= &\left(\frac{n_{B}}{n_{\phi}}\right)\left(
\frac{n_{\phi}}{s}\right)\, \nonumber \\
&\simeq &{1 \over \pi} {{\delta h} h'^{2} \over
h^3} {\Delta M_N \over M_N} \left({M_N m_{\nu} \over {\langle
H^{0}_{u}\rangle}^2}\right)
\times \left ({m_\phi \over M_N}\right )^2\left({T_{\rm R} \over
m_\phi}\right)\,,
\end{eqnarray}
where $s=(2\pi^2/45)g_{\ast}T_{\rm R}^3$. Here
$n_{\phi}/s$ denotes the dilution from reheating. By using
Eq.~(\ref{rehtemp}), and the relationship
$m_\nu \simeq (h^2 \langle H^{0}_{u}\rangle^2/ M_N)$, we eventually obtain
\beq 
\label{result}
\eta_{\rm B} \simeq 4.10^{-49/2} g {{\delta h} h'^{2} \over
h^3} {\Delta M_N \over M_N} {m^{7/2}_{\phi}
M^{1/2}_{\rm
P} \over M^{2}_{N} {\langle H^{0}_{u} \rangle}^4}~~(1 {\rm GeV})^2,
\eeq        
where we have taken $m_\nu \approx 0.1$~eV. We also assume 
$\langle H^{0}_{u} \rangle = 174$ GeV in below. Moreover, 
for ${\Delta M}_{N} \simeq M_N$ and as long as $h' < \delta h$, 
it is sufficient to have ${\delta h}/h \approx \Delta M_N/2 M_N$ 
in order to obtain degenerate light neutrino masses. Therefore, we 
may further simplify Eq. (\ref{result}) to find
\beq 
\label{finalresult}
\eta_{\rm B} \simeq 2.10^{-49/2} g {h'^{2} \over
h^2} \left({\Delta M_N \over M_N}\right)^2 {m^{7/2}_{\phi}
M^{1/2}_{\rm
P} \over M^{2}_{N} {\langle H^{0}_{u} \rangle}^4}~~(1 {\rm GeV})^2. 
\eeq
 
Let us now present some numerical examples for nearly degenerate
superheavy RH (s)neutrinos, i.e. 
$M_N \geq 10 m_\phi$ and $\Delta M_N \simeq M_N$. With $M_N = 10 m_\phi$ 
and $10^{-1} \leq h'/h \leq 1$, the desired baryon asymmetry can be
obtained for the range of parameters $10^{-3}\leq g \leq 1$ and 
$10^{11}~{\rm GeV} \leq m_\phi \leq 10^{13}$~GeV, which result 
in $10^6~{\rm GeV} \leq T_{\rm R} \leq 10^8$~GeV.
With $M_N = 100 m_\phi$, and $10^{-1} \leq h'/h \leq 1$ as before, an
acceptable asymmetry is yielded for $g = 1$ and 
$m_\phi \simeq 10^{12}-10^{13}$ GeV, with 
$10^{7}~{\rm GeV} \leq T_{\rm R} \leq 10^{9}$ GeV.

The merits of our model are already evident from these numbers. 
First of all, the reheat temperature is low (more than) enough to
avoid the gravitino problem. Moreover, $T_{\rm R} \ll M_N$ guarantees
that lepton number violating scattering of the SM particles are 
completely negligible. Especially, keeping in mind that in the 
MSSM, there are a large number of scattering processes which can 
considerably attenuate the obtained asymmetry if the reheat 
temperature $T_{\rm R}$ is close to $M_N$ \cite{plumacher98}. 
In our case, obtaining a sufficiently low reheat temperature is 
more than welcome in this regard. Also, the robustness of the
inflaton mass $m_{\phi}$ lies in a range compatible with both high and
intermediate scale inflationary models, though slightly favoring high
scale models, thus making the scenario more flexible.

We shall re-emphasize the marked difference from leptogenesis with
on-shell (s)neutrinos, namely suppression of the yielded asymmetry as
$\Delta M_N/M_N$ decreases. This implies that our scenario works well
for nearly degenerate neutrinos (and perhaps even better in the
hierarchical case), while producing too little asymmetry for highly
degenerate ones. Note that no resonant enhancement of the 
lepton asymmetry of the type discussed in Ref. \cite{pilaftsis} will occur. 
However, we can expect a qualitatively similar effect if (at least) one of 
the RH sneutrinos is almost degenerate with the inflaton.

And the final comment before closing this section. A small number of
on-shell (s)neutrinos might also have been produced non-perturbatively, 
from an inefficient preheating, and hence contribute to the resultant 
asymmetry through their decay. The asymmetry yielded in the decay of on-shell
particles, denoted as $\eta^{on}_{\rm B}$, will be
\beq 
\label{on-shell}
\eta^{on}_{\rm B} \simeq {M^{4}_{N} \over 3 {\Delta M_N}^{2}
m^{2}_{\phi}}
\left({n_{\tilde N} + n_N \over n_\phi}\right) \eta_{\rm B}.
\eeq
Note that the asymmetry parameter for on-shell
(s)neutrinos is dominated by the self-energy correction, given in
Eq. (\ref{corrections}), and hence $\eta^{on}_{\rm B}$ does not
contain the suppression factor $\left(m_\phi/2 M_N \right)^2$. 
On the other hand, a factor of $4$ will be lost, relative to the 
off-shell case, since the one-particle decay of on-shell $\tilde N$ and $N$
violate the lepton number by one unit. Thus, with $\Delta M_N \simeq M_N$, 
possible contribution from on-shell (s)neutrinos can be neglected, provided 
$(n_{\tilde N} + n_N) < \left(3 m^{2}_{\phi}/M^{2}_N \right) n_\phi$. 
For the range of parameters considered above this is generically
the case.

\section{conclusion}

In this paper we have provided a simple example for non-thermal 
leptogenesis with nearly degenerate superheavy RH neutrinos in 
a supersymmetric set up. We assumed that the inflaton is lighter 
than the RH (s)neutrinos, thus naturally avoiding some potential
problems which can naturally arise. The inflaton decay via off-shell
(s)neutrinos reheats the Universe and the model is minimal in a 
sense that the same channel is also responsible for generating the 
lepton asymmetry. As usual, the asymmetry arises from the interference 
between the tree-level and the one-loop diagrams representing self-energy 
and vertex corrections of (s)neutrinos, although off-shell in our
case, provided neutrino Yukawa couplings contain $CP$-violating
phases. However, there are important differences from leptogenesis 
with on-shell (s)neutrinos, which we have pronounced here. The
self-energy and vertex corrections are now of the same order
regardless of the degree of degeneracy. Most notably, the asymmetry
parameter is found to be linearly proportional (rather than inversely
in the on-shell case) to the mass difference of the RH (s)neutrinos. 
This results in a lepton asymmetry which gets smaller as the RH
(s)neutrinos become more degenerate.

Finally, we briefly emphasize on remarkable advantages of this
model. First of all leptogenesis can be accommodated rather simply
without relying on non-perturbative production of RH (s)neutrinos.
It is particularly attractive that the desired baryon asymmetry can
be directly generated in the final stage of reheating which is
perturbative, regardless of any model-dependent effects which 
might have resulted in a first stage of non-perturbative reheating. 
Secondly, the suppressed decay of the inflaton naturally leads to 
an acceptably low reheat temperature, which is compatible with the 
gravitino bound and also prevents any wash out of the yielded 
asymmetry. Also, with nearly degenerate (s)neutrinos, the desired 
lepton asymmetry can be generated for a range of inflaton mass 
accessible in large and intermediate scale models of inflation.

Qualitatively, we expect that this scenario also work (even
better) in the case of hierarchical RH (s)neutrinos. However, a
more careful study should be performed in order to compare the 
quantitative results with those obtained here. It will also be interesting to 
study possible enhancement of the lepton asymmetry when the inflaton is 
almost degenerate with some of the RH sneutrinos.


\acknowledgements
We wish to thank Z. Berezhiani, W. B\"uchmuller, M. Drees, B. Dutta,
and A. Perez-Lorenzana for useful discussions. The work of R.A. was 
supported by ``Sonderforschungsbereich 375 f\"ur
Astro-Teilchenphysik'' der
Deutschen Forschungsgemeinschaft. A.M. acknowledges the support of
{\it The Early Universe Network} HPRN-CT-2000-00152.  


\end{document}